\documentclass{PoS}
\usepackage{graphicx}
\usepackage{subcaption}
\usepackage{amsmath}
\usepackage{braket}
\usepackage{tikz}

\title{Structure of hybrid static potential flux tubes in lattice Yang-Mills theory}

\ShortTitle{Structure of hybrid static potential flux tubes}

\author{\speaker{Lasse Mueller}\\
	Goethe University Frankfurt\\
	E-mail: \email{lmueller@th.physik.uni-frankfurt.de}}

\author{Owe Philipsen\\
	Goethe University Frankfurt\\
	E-mail: \email{philipsen@th.physik.uni-frankfurt.de}}
    
\author{Christian Reisinger\\
	Goethe University Frankfurt\\
	E-mail: \email{reisinger@th.physik.uni-frankfurt.de}}

\author{Marc Wagner\\
	Goethe University Frankfurt\\
	E-mail: \email{mwagner@th.physik.uni-frankfurt.de}}

\abstract{We report about an ongoing lattice field theory project concerned with the investigation of heavy hybrid mesons. In particular we discuss our computation of the structure of hybrid static potential flux tubes in SU(2) lattice Yang-Mills theory, which is based on the squares of the chromoelectric and chromomagnetic field strength components. Our flux tube results for hybrid static potential quantum numbers $\Sigma_g^-$, $\Sigma_u^+$, $\Sigma_u^-$, $\Pi_g$, $\Pi_u$, $\Delta_g$, $\Delta_u$ are significantly different compared to the flux tube of the ordinary static potential.}

\FullConference{XIII Quark Confinement and the Hadron Spectrum - Confinement2018\\
		31 July - 6 August 2018\\
		Maynooth University, Ireland}

\begin{document}


\section{Introduction}

Quark-antiquark systems with excited gluons are called hybrid mesons. Hybrid mesons are not limited to quark model quantum numbers $J^{\mathcal{P} \mathcal{C}}$ with $\mathcal{P} = (-1)^{L+1}$ and $\mathcal{C} = (-1)^{L+S}$, where $L \in \{ 0,1,2,\ldots \}$ is the orbital angular momentum and $S \in \{ 0,1 \}$ is the quark spin. Hybrid mesons are and will be investigated by ongoing and future experiments, e.g.\ the GlueX and the PANDA experiments (for a review cf.\ \cite{Olsen:2017bmm}. Also on the theoretical side there is currently a lot of activity, to improve our understanding of hybrid mesons and other exotic hadrons (for a review cf.\ \cite{Meyer:2015eta}).

A common approach in lattice gauge theory to study hybrid mesons is to compute hybrid static potentials and related quantities, e.g.\ the gluonic distribution around a static quark-antiquark pair with non-trivial quantum numbers. For existing lattice work on hybrid static potentials cf.\ e.g.\ \cite{Juge:1997nc,Michael:1998tr,Bali:2000vr,
Juge:2002br,Michael:2003ai,Bali:2003jq,
Wolf:2014tta,Reisinger:2017btr}. Very recently also hybrid static potential flux tubes have been computed \cite{Bicudo:2018yhk,Mueller:2018fkg,Bicudo:2018jbb}. Lattice gauge theory results on hybrid static potentials are often used as input in effective field theory or phenomenological approaches, e.g.\ to predict the spectrum of heavy hybrid mesons in the Born-Oppenheimer approximation (cf.\ e.g.\ \cite{Braaten:2014qka,Berwein:2015vca,Oncala:2017hop,
Brambilla:2017uyf,Brambilla:2018pyn}).


\section{Theoretical basics of the computation of hybrid static potentials and flux tubes}

We study hybrid mesons with a static quark $Q$ and a static antiquark $\bar{Q}$, which are treated as spinless color charges. Since their positions are frozen, symmetries and quantum numbers are different from the usual classification via $J^{\mathcal{P} \mathcal{C}}$. There are three quantum numbers (for a more detailed discussion cf.\ e.g.\ \cite{Bali:2005fu,Bicudo:2015kna}).
\begin{itemize}
\item The absolute value of total angular momentum with respect to the $Q \bar{Q}$ separation axis (in this work the $z$ axis): $\Lambda \in \{ \Sigma \doteq 0 , \Pi \doteq 1 , \Delta \doteq 2 , \ldots \}$.

\item The eigenvalue of parity combined with charge conjugation $\mathcal{P} \circ \mathcal{C}$: $\eta \in \{ g \doteq + , u \doteq - \}$.

\item The eigenvalue of the spatial reflection along an axis perpendicular to the $Q \bar{Q}$ separation axis $\mathcal{P}_x$ (in this work the $x$ axis): $\epsilon \in \{ +, - \}$.
\end{itemize}

After computing ordinary rectangular Wilson loops $W(r,t)$ with spatial extents $r$ (corresponding to the $Q \bar{Q}$ separation) and temporal extents $t$ on an ensemble of gauge link configurations, the ordinary static potential with quantum numbers $\Lambda_\eta^\epsilon = \Sigma_g^+$ can be extracted using
\begin{align}
V_{\Sigma_g^+}(r) = \lim_{t \rightarrow \infty} V_{\Sigma_g^+,\textrm{eff}}(r,t) \quad , \quad V_{\Sigma_g^+,\textrm{eff}}(r,t) = \frac{1}{a} \ln\bigg(\frac{\braket{W(r,t)}}{W(r,t+a)}\bigg) .
\end{align}
In practice one typically fits a constant in a plateau-like region at larger values of $t$ for given $r$ to obtain $V_{\Sigma_g^+}(r)$.

\begin{figure}[htb]
		\centering
		\captionsetup[subfigure]{labelformat=empty}
		\begin{subfigure}[b]{0.25\textwidth}
			\includegraphics[width=\textwidth]{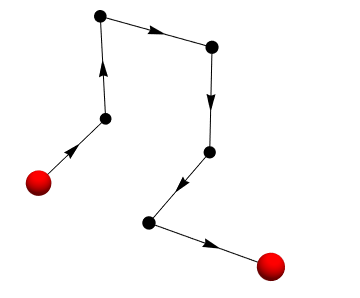}
			\caption{$\Sigma_g^-$ , $\Sigma_u^+$}
		\end{subfigure}
		\begin{subfigure}[b]{0.25\textwidth}
			\includegraphics[width=\textwidth]{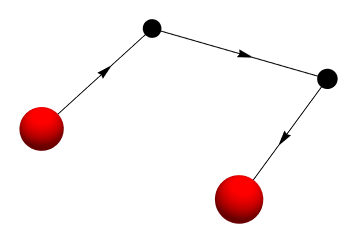}
			\caption{$\Delta_g$ , $\Pi_u$}
		\end{subfigure}
		\\
		\begin{subfigure}[b]{0.3\textwidth}
			\includegraphics[width=\textwidth]{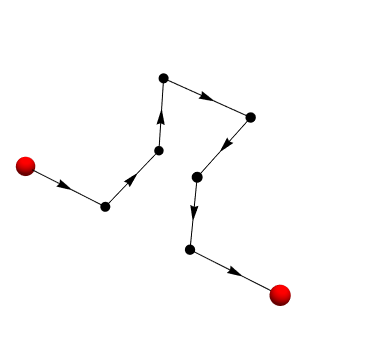}
			\caption{$\Sigma_u^-$}
		\end{subfigure}
		\begin{subfigure}[b]{0.3\textwidth}
			\includegraphics[width=\textwidth]{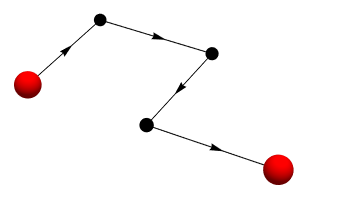}
			\caption{$\Pi_g$}
		\end{subfigure}
		\begin{subfigure}[b]{0.3\textwidth}
			\includegraphics[width=\textwidth]{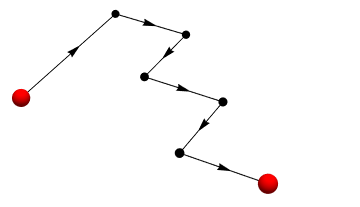}
			\caption{$\Delta_u$}
		\end{subfigure}
		\caption{\label{fig:optimized_operators}Operators $O_S$ used to construct hybrid static potential trial states with quantum numbers $\Lambda_\eta^\epsilon$ via eq.\ (\ref{EQN001}) for the flux tube computations.}
	\end{figure}

The computation of hybrid static potentials $V_{\Lambda_\eta^\epsilon}(r)$ is quite similar. One has to replace the spatial parts of an ordinary Wilson loop, which are straight lines of gauge links, by more complex structures, which generate quantum numbers $\Lambda_\eta^\epsilon$ different from $\Sigma_g^+$. A general prescription to construct a trial state with quantum numbers $\Lambda^\eta_\epsilon$ is given by
\begin{align}
\label{EQN001} \ket{\Psi_{\textrm{hybrid},\Lambda_\eta^\epsilon}} = \frac{1}{4}
\Big(1 + \eta (\mathcal{P} \circ \mathcal{C}) + \epsilon \mathcal{P}_x + \eta \epsilon (\mathcal{P}\circ \mathcal{C}) \mathcal{P}_x\Big)
\sum_{k=0}^{3} \exp\bigg(\frac{i \pi \Lambda k}{2}\bigg) R\bigg(\frac{\pi k}{2}\bigg) O_S \ket{\Omega} .
\end{align}
$R$ denotes a rotation around the $Q \bar{Q}$ separation axis and $O_S$ is a non-straight path of spatial gauge links connecting the quark and the antiquark. Example operators $O_S$, which we use to compute the flux tubes, are shown in Figure~\ref{fig:optimized_operators}. The computation of the potentials is done with a highly optimized and very large set of different operators, which is discussed in detail in a very recent publication \cite{Reisinger:2018lne}.

To get insights on the gluonic distribution inside hybrid mesons we compute the squares of the chromoelectric and chromomagnetic field strength components in presence of a static quark-antiquark pair with quantum numbers $\Lambda_\eta^\epsilon$ and subtract the corresponding vacuum expectation values,
\begin{align}
\label{Eqn: E_squared} \Delta E_j^2 \equiv \braket{E_j(\mathbf{x})^2}_{Q\bar{Q}} - \braket{E_j^2}_{\textrm{vac}}
& \propto \bigg(\frac{\braket{W \cdot P_{0 j}(t/2, \mathbf{x})}}{\braket{W}} - \braket{P_{0 j}} \bigg) \\
\label{Eqn: B_squared} \Delta B_j^2 \equiv \braket{B_j(\mathbf{x})^2}_{Q\bar{Q}} - \braket{B_j^2}_{\textrm{vac}}
& \propto \bigg(\braket{P_{k l}} - \frac{\braket{W \cdot P_{k l}(t/2, \mathbf{x})}}{\braket{W}} \bigg) , 
\end{align}
where $j,k,l \in \{1,2,3\}$, $W$ denotes a generalized Wilson loop (i.e.\ spatial parts replaced by linear combinations of $O_S$ according to eq.\ (\ref{EQN001})) and $P_{\mu\nu}$ the plaquette.


\section{Numerical results for hybrid static potentials}

Computations have been performed on $5 \, 500$ SU(3) gauge link configurations generated with the Wilson plaquette action using the Chroma QCD library \cite{Edwards:2004sx}. The lattice volume is $48 \times 24^3$. We have used gauge coupling $\beta = 6.0$, which corresponds to lattice spacing $a \approx 0.093 \, \textrm{fm}$, when identifying $r_0$ with $0.5 \, \textrm{fm}$. To improve the quality of the signal, temporal links are HYP2 smeared, while spatial links are APE smeared with parameters tuned to optimize the ground state overlaps.

\begin{figure}[htb]
		\centering
    \includegraphics[width=12.0cm]{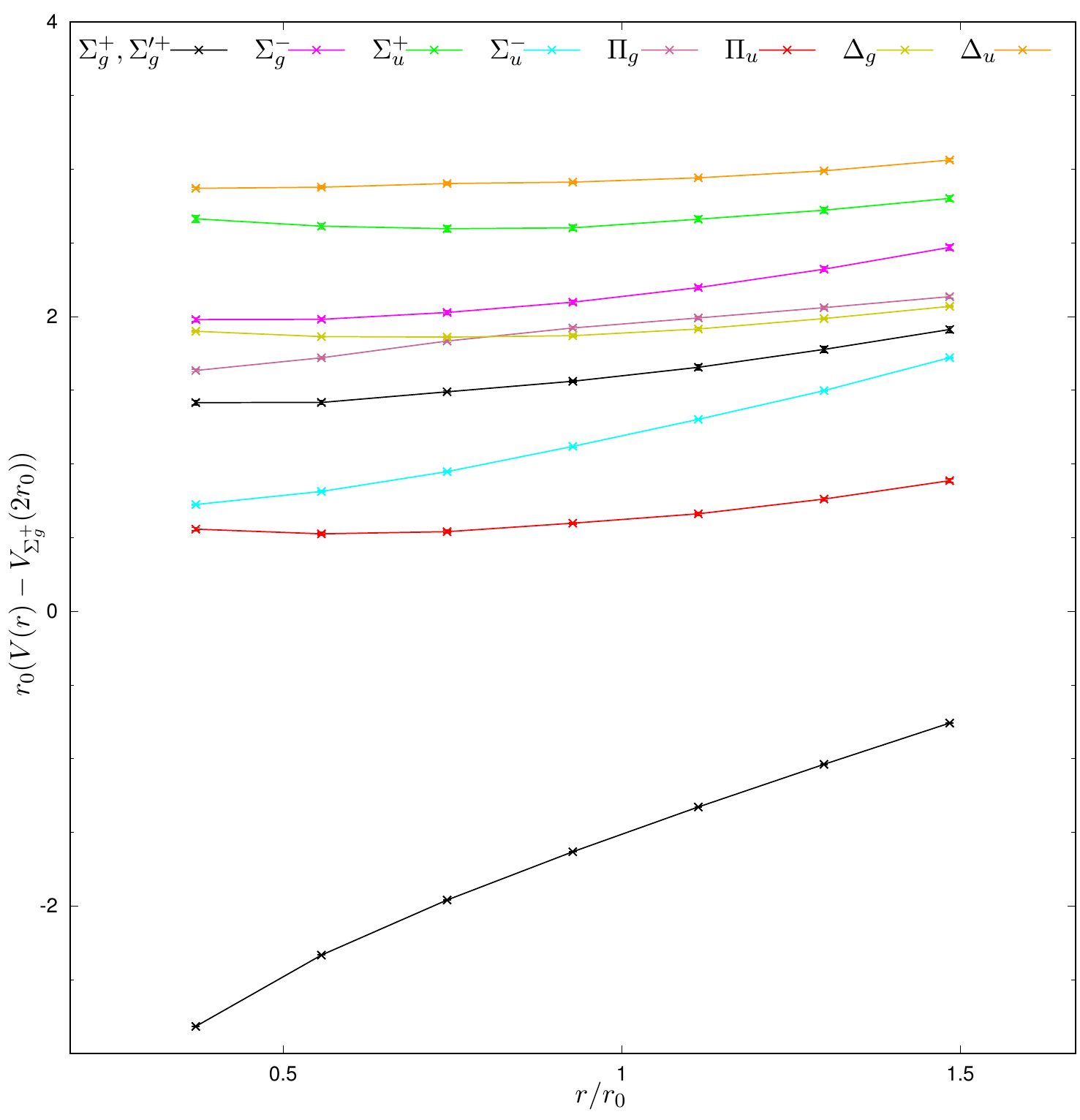}
		\caption{\label{fig:potential}The ordinary static potential ($\Sigma_g^+$) and its first excitation ($\Sigma'^+_g$) as well as the hybrid static potentials with quantum numbers $\Sigma_g^-$, $\Sigma_u^+$, $\Sigma_u^-$, $\Pi_g$, $\Pi_u$, $\Delta_g$, $\Delta_u$ in units of $r_0 = 0.5 \, \textrm{fm}$.}
	\end{figure}

The resulting ordinary static potential ($\Sigma_g^+$) and its first excitation ($\Sigma'^+_g$) as well as the hybrid static potentials with quantum numbers $\Sigma_g^-$, $\Sigma_u^+$, $\Sigma_u^-$, $\Pi_g$, $\Pi_u$, $\Delta_g$, $\Delta_u$ are shown in Figure~\ref{fig:potential}. They are plotted in units of $r_0$ to allow a straightforward comparison to results from the literature, in particular to \cite{Juge:1997nc,Juge:2002br}, which are frequently used in recent publications (cf.\ e.g.\ \cite{Berwein:2015vca,Oncala:2017hop}) and seem to be the most accurate lattice results for hybrid static potentials, which are currently available. A detailed discussion is given in a very recent publication \cite{Reisinger:2018lne}, where we also provide numerical values for all data points.

\section{Numerical results for hybrid static potential flux tubes}

Computations have been performed on $13 \, 000$ SU(2) gauge link configurations generated with the Wilson plaquette action. The lattice volume is $18^4$. We have used gauge coupling $\beta = 2.5$, which corresponds to lattice spacing $a \approx 0.073 \, \textrm{fm}$, when identifying $r_0$ with $0.5 \, \textrm{fm}$. The discretization of $E_j^2$ and $B_j^2$ at a given spacetime point is done by averaging over the four adjacent and correspondingly oriented plaquettes, i.e.\ by clover leaf-like combinations of gauge links. As before, spatial links are APE smeared with parameters tuned to optimize the ground state overlaps.

\begin{figure}[htb]
		\centering
		\includegraphics[width = \textwidth, trim = 0cm 0.5cm 0cm 0cm]{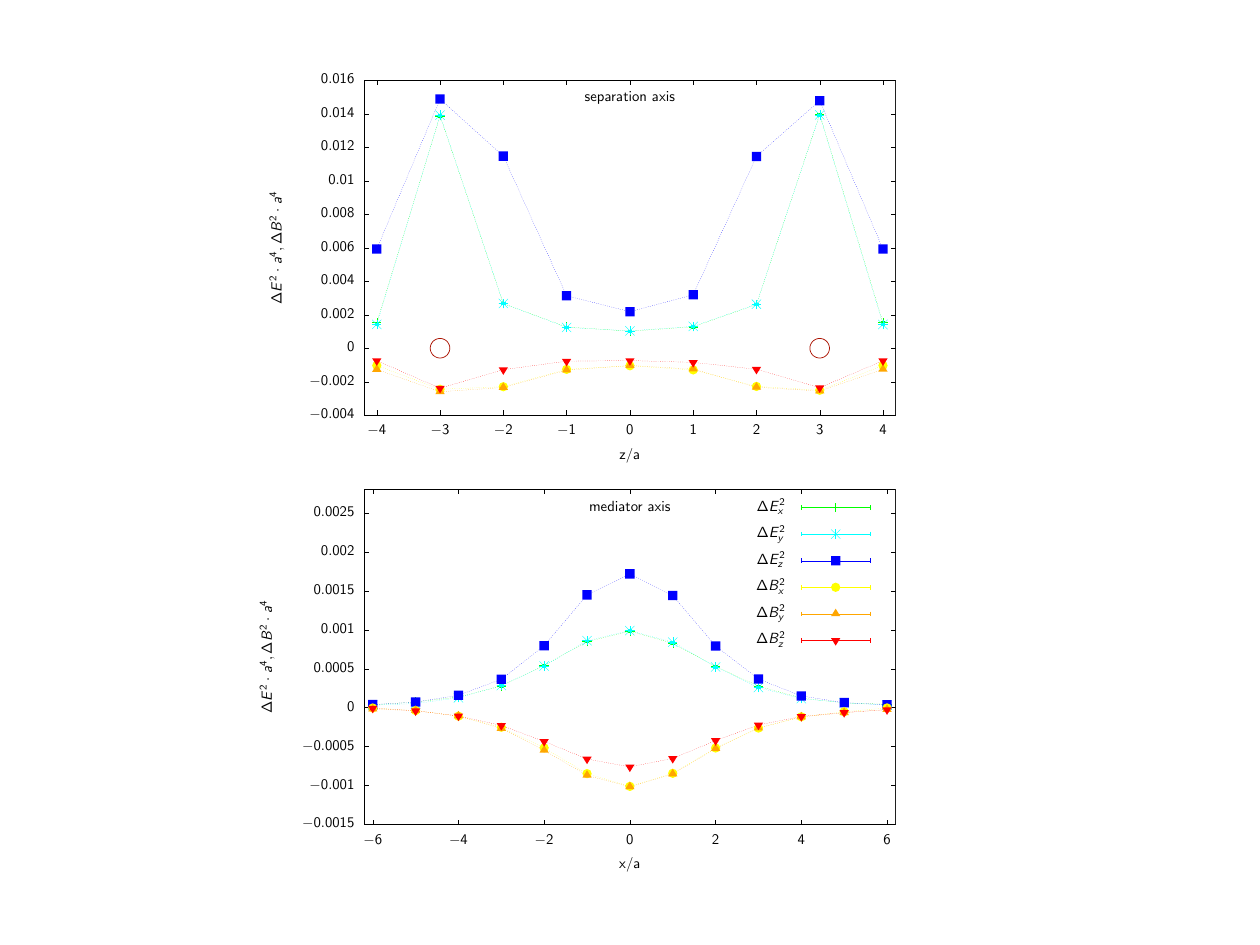}
		\caption{\label{fig:ordinary_flux_tubes}$\Delta E_j^2$ and $\Delta B_j^2$ for the ordinary static potential ($\Lambda_\eta^\epsilon = \Sigma_g^+$) on the separation axis and the mediator axis.}
	\end{figure}

In the following results a rather small temporal separation of $t = 2$ was used to get a favorable signal-to-noise ratio. This leads to results that contain contributions from the first excited states in addition to the ground state. Results using larger separations $t$ do not show discrepancies within their errors, confirming shown results qualitatively. \\
In Figure~\ref{fig:ordinary_flux_tubes} we show the squared field strength components $\Delta E^2$ and $\Delta B^2$ (eqs.\ (\ref{Eqn: E_squared}) and (\ref{Eqn: B_squared})) for the ordinary static potential ($\Lambda_\eta^\epsilon = \Sigma_g^+$) on the $Q \bar{Q}$ separation axis and the mediator axis, i.e.\ an axis perpendicular to the separation axis centered between the quark and the antiquark. The results on the separation axis are generated without HYP2 smeared temporal links, because such a smearing enhances lattice discretization errors in the vicinity of the static quarks. For the results on the mediator axis HYP2 smeared temporal links are used.

\begin{figure}[htb]
		\centering
		\includegraphics[width = \textwidth, trim = 0cm 0.5cm 0cm 0cm]{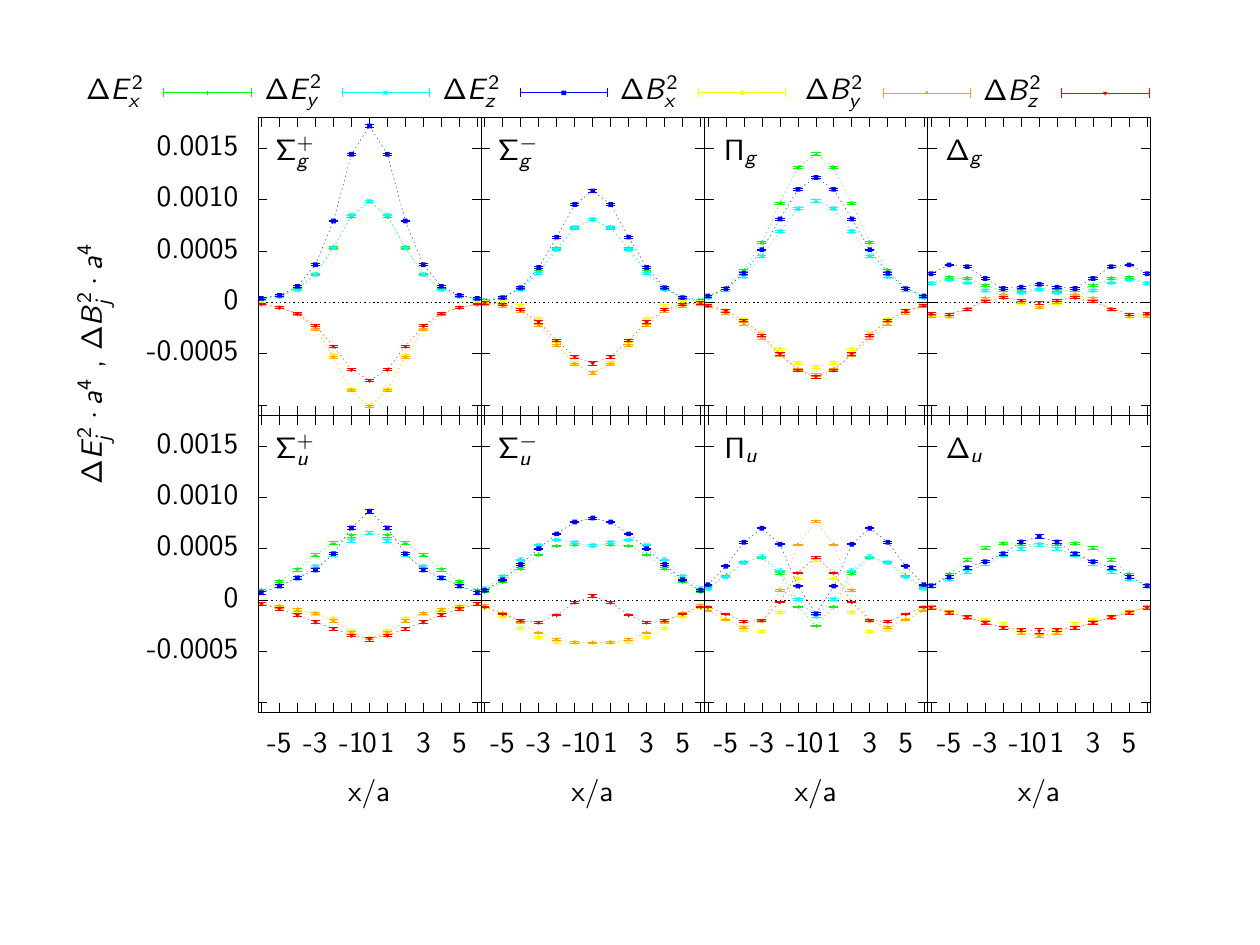}
		\caption{\label{fig:mediator_axis_hybrid}$\Delta E_j^2$ and $\Delta B_j^2$ for the ordinary static potential ($\Lambda_\eta^\epsilon = \Sigma_g^+$)and the hybrid static potentials with $\Lambda_\eta^\epsilon = \Sigma_g^- , \Sigma_u^+ , \Sigma_u^- , \Pi_g , \Pi_u , \Delta_g , \Delta_u$ on the mediator axis.} 
	\end{figure}

As already observed in \cite{Mueller:2018fkg}, the field strength components of the ordinary and of hybrid static potentials are essentially the same in the vicinity of the static quarks. Therefore, we focus on results on the mediator axis. In Figure~\ref{fig:mediator_axis_hybrid} we compare the squared field strength components of the ordinary static potential ($\Lambda_\eta^\epsilon = \Sigma_g^+$) and the hybrid static potentials with $\Lambda_\eta^\epsilon = \Sigma_g^- , \Sigma_u^+ , \Sigma_u^- , \Pi_g , \Pi_u , \Delta_g , \Delta_u$. Clearly, there are significant qualitative differences between these flux tube profiles. For example, the $\Pi_u$ sector exhibits zero crossings for all six field strength components, while in the other sectors such zero crossings are not present. Another example is the $\Pi_g$ sector, where $\Delta E_x^2$ is dominating, while in many other sectors $\Delta E_z^2$ is dominating. Note that $\Delta E_x^2 = \Delta E_y^2$ and $\Delta E_x^2 = \Delta E_y^2$ only for $\Lambda = \Sigma$, as one can show analytically.

\begin{figure}[t]
		\centering
		\includegraphics[width = \textwidth, trim = 0cm 0.5cm 0cm 0cm]{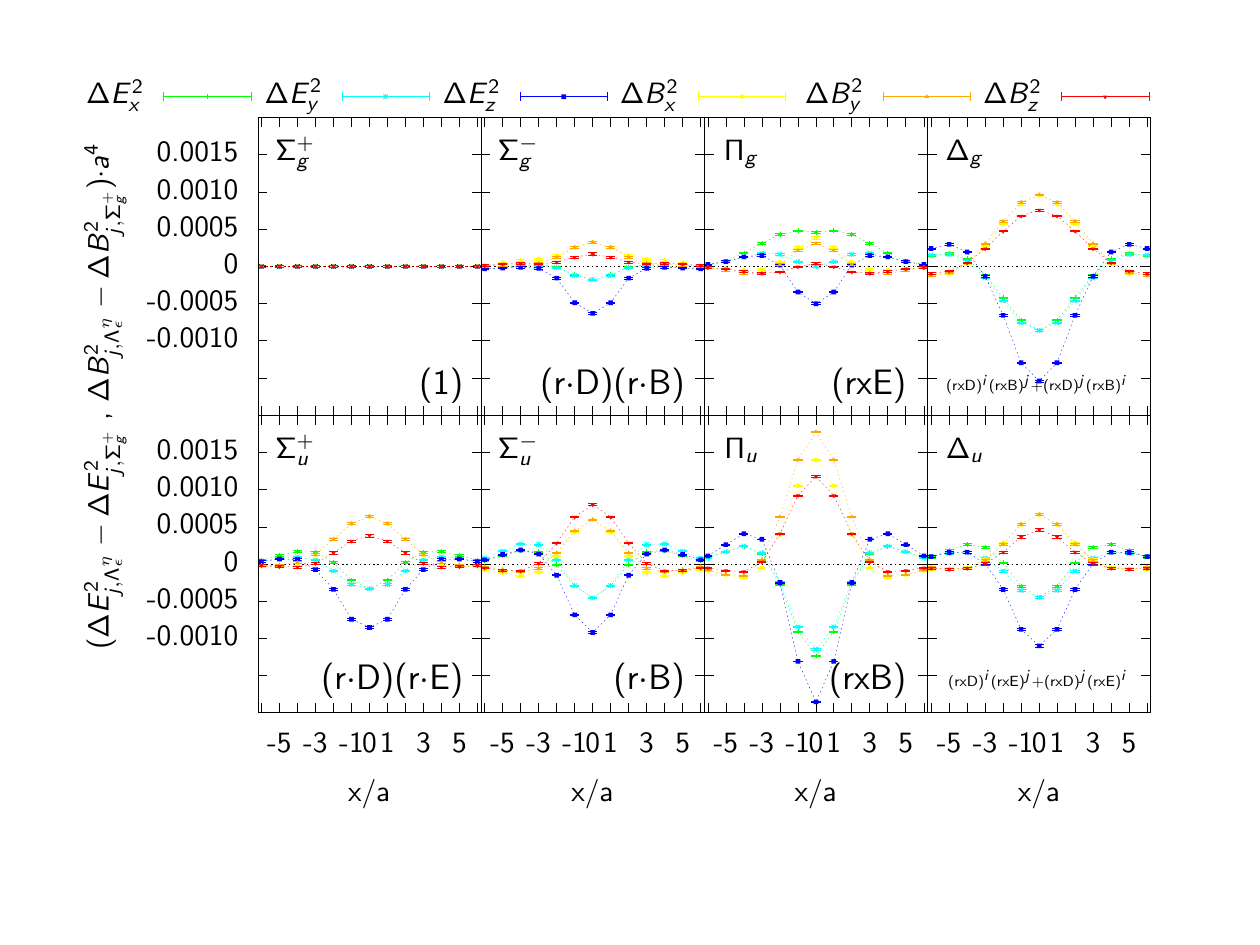}
		\caption{\label{fig:diff_mediator_axis_hybrid_with_operators}$\Delta E_{j,\Lambda_\eta^\epsilon}^2 - \Delta E_{j,\Sigma_g^+}^2$ and $\Delta B_{j,\Lambda_\eta^\epsilon}^2 - \Delta B_{j,\Sigma_g^+}^2$ for the hybrid static potentials with $\Lambda_\eta^\epsilon = \Sigma_g^- , \Sigma_u^+ , \Sigma_u^- , \Pi_g , \Pi_u , \Delta_g , \Delta_u$ on the mediator axis.}
	\end{figure}

To exhibit the difference of the hybrid static potential flux tubes to that of the ordinary static potential, we show $\Delta E_{j,\Lambda_\eta^\epsilon}^2 - \Delta E_{j,\Sigma_g^+}^2$ and $\Delta B_{j,\Lambda_\eta^\epsilon}^2 - \Delta B_{j,\Sigma_g^+}^2$ in Figure~\ref{fig:diff_mediator_axis_hybrid_with_operators}. One can see that the dominance of $\Delta E_z^2$ is much weaker for the hybrid static potentials than for the ordinary static potential. Moreover, the chromomagnetic field strength components are enhanced for the hybrid static potentials. We interpret this localization of chromomagnetic flux as gluons generating the hybrid quantum numbers. 

While we have found that it is numerically advantageous to use spatially extended creation operators (cf.\ e.g.\ Figure~\ref{fig:optimized_operators}), it is also possible and common to consider local operators, which are inserted at the center of a straight parallel transporter connecting the static quark and the static antiquark. An example is the pNRQCD investigation \cite{Berwein:2015vca}, where these local insertions are listed in Table~I (they are also included in Figure~\ref{fig:diff_mediator_axis_hybrid_with_operators} of this work). It is interesting to note that the field strength components we have computed show similarities to these local operators, in particular for $\Lambda_\eta^\epsilon = \Sigma_u^- , \Pi_g , \Pi_u$. For example for $\Lambda_\eta^\epsilon = \Sigma_u^-$ the local operator is $\hat{\mathbf{r}} \cdot \mathbf{B} = B_z$, which is consistent with our Figure~\ref{fig:diff_mediator_axis_hybrid_with_operators}, where the dominating squared field strength component is $\Delta B_z^2$


\section*{Acknowledgements}

L.M.\ thanks M.\ Faber and all organizers of the ``XIIIth Quark Confinement and the Hadron Spectrum'' conference for the invitation to give this talk.

We acknowledge useful discussions with P.\ Bicudo.

C.R.\ acknowledges support by a Karin and Carlo Giersch Scholarship of the Giersch foundation. O.P.\ and M.W.\ acknowledge support by the DFG (German Research Foundation), grants PH 158/4-1 and WA 3000/2-1. M.W.\ acknowledges support by the Emmy Noether Programme of the DFG, grant WA 3000/1-1.

This work was supported in part by the Helmholtz International Center for FAIR within the framework of the LOEWE program launched by the State of Hesse.

Calculations on the LOEWE-CSC and on the on the FUCHS-CSC high-performance computer of the Frankfurt University were conducted for this research. We would like to thank HPC-Hessen, funded by the State Ministry of Higher Education, Research and the Arts, for programming advice.


\end{document}